\documentclass{aa}  

\usepackage{graphicx}
\usepackage{txfonts}
\usepackage{newtxtext,newtxmath}
\usepackage[T1]{fontenc}
\DeclareRobustCommand{\VAN}[3]{#2}
\let\VANthebibliography\thebibliography
\def\thebibliography{\DeclareRobustCommand{\VAN}[3]{##3}\VANthebibliography}
\usepackage{graphicx}
\usepackage{amsmath}
\usepackage{hyperref}
\usepackage{subcaption} 
\usepackage{multirow}
\usepackage{makecell}
\usepackage{rotating}
\usepackage{pdflscape}
\usepackage{float}
\let\oldthebibliography\thebibliography
\let\endoldthebibliography\endthebibliography
\renewenvironment{thebibliography}[1]
  {\oldthebibliography{#1}
   \setlength{\itemsep}{2pt} 
   \setlength{\parskip}{1pt} 
   \setlength{\parsep}{1pt}}  
  {\endoldthebibliography}

\usepackage{caption}
\DeclareCaptionFormat{custom}
{
    \textbf{#1#2}{#3}
}
\DeclareCaptionLabelSeparator{custom}{. }
\begin{document}

   \title{MaNGA AGN dwarf galaxies (MAD) -- IV. Revealing hidden AGN in dwarf galaxies with radio observations}

   \author{I. Flores\inst{1}, M. Mezcua\inst{2, 3} \and V. Rodr\'{i}guez Morales\inst{2}}

   \institute{Donosti International Physics Center (DIPC), Manuel Lardizabal Ibilbidea, 4,                  20018, Donostia, Spain\\
              \email{irene.valderrama@dipc.org}
         \and
             Institute of Space Sciences (ICE, CSIC), Campus UAB, Carrer de Magrans, 08193, Barcelona, Spain
         \and
            Institut d'Estudis Espacials de Catalunya (IEEC),  Edifici RDIT, Campus UPC, 08860 Castelldefels, Barcelona, Spain
             }

   \date{Received XXX; accepted YYY}

  \abstract 
   {Low-mass black holes hosted by dwarf galaxies offer valuable insights into galaxy formation and the growth of the more massive black holes found in massive galaxies. Their detection as Active Galactic Nuclei (AGN) is challenging due to their low luminosity and compact size. This can be circumvented employing multi-wavelength observational strategies, such as combining optical and radio observations, which enables the detection of AGN features that may be hidden in single-wavelength analyses.}
   {We aim to detect any jet-like emission indicative of the presence of an AGN in a sample of four dwarf galaxies with AGN signatures based on spatially resolved emission line diagnostic diagrams with SDSS Mapping Nearby Galaxies at APO (MaNGA) optical integral-field  unit (IFU) spectroscopy. Confirming the presence of an AGN will prove IFU spectroscopy to be a resourceful tool for identifying hidden or switched-off AGN.}
   {Using Karl G. Jansky Very Large Array (VLA) radio observations, we image the radio emission of the four dwarf galaxies and derive their integrated radio flux and luminosity. We compare these to that expected from star formation processes to determine the origin of the radio emission and probe if it is consistent with the results of the emission line diagnostic diagrams.}
   {We find that one out of the four galaxies shows AGN radio emission consistent with the analysis of the MaNGA IFU data. The kinetic jet power of this source is $Q_{\mathrm{jet}} \sim 10^{42}$ erg s$^{-1}$, indicating that dwarf galaxies can host radio jets as powerful as those of massive radio galaxies, whose jet mechanical feedback can strongly affect the star formation in the host galaxy. Furthermore, this galaxy exhibits an AGN outflow able to escape the gravitational bound produced by the dark matter halo, along with a decrease in the star formation rate of the central region. This suggests the presence of negative feedback from the AGN, which could be suppressing star formation. The other three galaxies exhibit regions of radio emission consistent with a stellar origin and overlapping with the star-forming regions found in the IFU spectroscopy.}
{}

   \titlerunning{ }
   \authorrunning{Flores et al.}
   
   \maketitle

\section{Introduction}
     Supermassive Black Holes (SMBHs; $M_{\mathrm{BH}} > 10^6 M_{\odot}$) are the most massive type of black holes (BHs) and are found in the nuclei of most, if not all, massive galaxies \citep{Kormendy2013}. The finding of SMBHs at redshift $z \sim 6 - 7 $ (e.g., \citeauthor{Fan2006} \citeyear{Fan2006}, \citeauthor{Mortlock2011} \citeyear{Mortlock2011}) and more recently at $z \sim 8 - 10$ (e.g., \citeauthor{Kokorev2023} \citeyear{Kokorev2023}, \citeauthor{Larson2023} \citeyear{Larson2023}, \citeauthor{Furtak2024} \citeyear{Furtak2024}, \citeauthor{Maiolino2024} \citeyear{Maiolino2024}) thanks to the James Web Space Telescope (JWST) would imply an alternative origin to that of stellar BHs ($M_{\mathrm{BH}} < 10^2$ M$_{\odot}$). A stellar BH seed would need more than $0.5$ Gyr to reach a mass of $10^9M_{\odot}$ accreting at the Eddington rate assuming a typical radiative efficiency of 10$\%$ \citep{Volonteri2010}. One possible explanation that accounts for such high masses at early times is that SMBHs could have been generated through accretion and mergers from $M_{\mathrm{BH}} > 10^2$ M$_{\odot}$ seeds created in the early Universe.  A detailed description of the different seed formation scenarios can be found in \citet{Mezcua2017} and \citet{Greene2020}. 
    \\
    BH seeds are thought to have formed during the early epochs of cosmic history, likely at high redshifts, making their direct detection and observation challenging. However, they can still be identified in the local Universe, particularly in low-mass Active Galactic Nuclei (AGN) within dwarf galaxies, where BH masses are typically below $10^6$ M$_{\odot}$ and stellar masses are $\mathrm{M}_* \leq 10^{10}$ M$_{\odot}$
    \citep{Mezcua2017,Reines2022}. These dwarf galaxies haven't grown much via mergers and accretion so are thought to resemble the first galaxies formed in the early universe \citep{Mezcua2017,Greene2020,Reines2022}. This makes them ideal environments for preserving BH seeds. 
    \\
    The AGN activity within dwarf galaxies can be detected across various regions of the electromagnetic spectrum. 
    In the optical, Balmer and forbidden line analysis helps identify AGN activity, with broad Balmer lines providing BH mass estimates. The Baldwin-Phillips-Terlevich diagram (BPT diagram after \citealt{Baldwin1981}) is a diagnostic tool based on [O III]/H$_\beta$, [N II]/H$_\alpha$, or [O I]/H$_\alpha$ emission line ratios used to differentiate the source of ionised gas (star formation, AGN, LINER\footnote{Low Ionization Emission Line Region.} or a combination of both; \citealt{kewley2001theoretical,kewley2006host}) that has been often applied to dwarf galaxies (e.g., \citealt{Reines2013}; \citealt{Mezcua2020,Mezcua2024}; \citealt{Pucha2024}). 
    
    However, traditional single-fibre surveys like Sloan Digital Sky Survey (SDSS; \citeauthor{Kauffmann2003} \citeyear{Kauffmann2003}) may miss low-mass AGN in star-forming dwarf galaxies \citep{Moran2002, Cann2019, Birchall2020} if other mechanisms rather than the AGN dominate the ionisation at the galaxy centre \citep{Moran2002}, if the AGN is displaced from the central region due to a galaxy merger \citep{Comerford2014, Barrows2018} or if the AGN has recently switched off, in which case its ionisation signatures are only observable as light echos at large distance from the centre \citep{Keel2015}. Integral Field Unit (IFU) spectroscopy, such as MaNGA (\citeauthor{Bundy2015} \citeyear{Bundy2015}), can circumvent this by obtaining multiple spectra across the field of view of the galaxy.

    Alternative ways to detect AGN that may be missed in single-fibre spectroscopy is via X-ray, mid-infrared, or radio observations. In X-rays, AGN are primarily detected through emissions from the accretion disk and hot corona surrounding the BH. These emissions are crucial for identifying faint AGN with low accretion rates, even at intermediate redshifts \citep{Mezcua2016, Mezcua2018}. However, X-ray signals can be ambiguous, as they may originate from X-ray binaries, which have similar luminosities to low-mass AGN. Moreover, detecting low-luminosity AGN requires long exposure times, adding to the complexity of this method. Mid-infrared emissions provide another means to detect AGN \citep{Lacy2004, Lacy2013, Hainline2014}. These emissions are generated when dust heated by the accreting BH reprocesses light, resulting in a characteristic infrared power-law spectrum. This method is less affected by obscuration from the host galaxy, but in dwarf galaxies, the low metallicity of the interstellar medium can lead to high temperatures and excitation levels that mimic AGN signatures \citep{Hainline2016}, complicating the identification of true AGN activity. Lastly, radio emission is produced in around 10$\%$ of AGN \citep{Ivezi2002} and originates primarily from the kinetic jets produced by the BH. Since radio wavelengths are unaffected by dust extinction, they provide an unobstructed view of AGN activity. Studies show that radio observations are crucial in understanding and detecting AGN feedback mechanisms (e.g. \citeauthor{Mezcua2019} \citeyear{Mezcua2019}, \citeauthor{Reines2020} \citeyear{Reines2020}, \citeauthor{Yang2023} \citeyear{Yang2023}, \citeauthor{Wu2024} \citeyear{Wu2024}, \citeauthor{Yuan2025} \citeyear{Yuan2025}). These jets not only influence BH accretion but also play a significant role in either quenching or triggering star formation in massive galaxies, thus affecting the host galaxy's evolution. Outflows coming from winds produced by the radiation pressure or from the accretion disk can also affect the star formation when interacting with the interstellar medium (see \citealt{harrison2024observational} for a recent review). Outflows can be identified by looking at complex profiles with broad wings, indicative of an extra kinematic component, in emission lines like [OIII]$\lambda$5007 or [NII]$\lambda$6583 (e.g. \citealt{harrison2014kiloparsec}; \citealt{leung2019mosdef}; \citealt{wylezalek2020ionized}; \citealt{rodriguez2025manga}). In the context of dwarf galaxies, outflows coming from supernovae were assumed to be the main mechanism that regulates the star formation. However, in the last decade, many studies provide evidence for AGN-driven outflows and feedback in dwarf galaxies (e.g. \citealt{penny2018sdss};\citealt{manzano2019agn, manzano2020active}; \citealt{liu2020integral, liu2024fast}; \citealt{zheng2023escaping}; \citealt{wang2024rubies}; \citealt{salehirad2025ionized}; \citealt{rodriguez2025manga}). By combining data from different wavelengths, a more complete picture of AGN activity and its impact on galaxies can be gained, especially in the challenging environments of dwarf galaxies (e.g. \citealt{molina2021outflows}, \citealt{Jin2025}).
    \\
    In this project, we examine data from four dwarf galaxies without AGN signatures in the optical SDSS spectrum, but with AGN signatures in the optical MaNGA IFU spectrum. To test whether these galaxies host an AGN, we perform a multi-wavelength analysis using radio observations with the Karl G. Jansky Very Large Array (VLA). We aim to resolve out the diffuse star formation emission and detect any jet-like emission supporting the presence of an AGN and, if feasible, derive the jet kinetic power from the radio luminosity of the compact core. This may help to find out whether dwarf galaxies host a radio jet as powerful as those of massive galaxies, whose jet mechanical feedback strongly affects the star formation of the host galaxy. Furthermore, for the dwarf galaxies  showing jet emission, we also investigate whether they have ionised AGN outflows based on the presence of a broad component in the [OIII]$\lambda$5007 emission line. This can have important implications for population studies in dwarf galaxies as well as for seed BH formation models.
    \\
    This paper is structured as follows: the properties of the targets and details of the VLA data reduction are presented in Section 2, the results and discussion are reported in Section 3, and finally conclusions are provided in Section 4. Throughout the paper we adopt a $\Lambda$CDM cosmology with parameters $\mathrm{H}_0 = 73$ km s$^{-1}$ Mpc$^{-1}$, $\Omega_\Lambda$ = 0.73 and  $\Omega_m$= 0.27.

\section{Sample and analysis}
\subsection{VLA data reduction}

    The four galaxies we analyse were selected from a sample of 23 newly identified AGN in star-forming dwarf galaxies by \citet{Mezcua2020} using MaNGA IFU data. The AGN signatures of these 23 dwarf galaxies are either hidden, partly hidden (in the case of the composite objects), or not available in the BPT of the SDSS spectra. From this sample, our four galaxies were chosen because they have radio counterparts in the Faint Images of the Radio Sky at Twenty-Centimeters (FIRST) survey at 1.4 GHz \citep{Becker1994}, which has a resolution of 5 arcseconds. Distinguishing whether this emission originates from AGN activity or star formation required follow-up observations with the VLA (project code 20B-023; PI: Mezcua). The VLA offers higher resolution and sensitivity, allowing us to resolve the extended radio emission detected by FIRST.   
    
    Our target sources are at $z < 0.04$, their stellar masses range from 
    $10^{9.7}$ M$_\odot$ to $10^{10.0}$ M$_\odot$ and their bolometric luminosities based on MaNGA IFU range from 39.2 to 40.2 erg s$^{-1}$.
    The sources were observed with the VLA at the C band, covering a bandwidth of 4 GHz centred at 6 GHz. The observations were taken on 2020 December 16. All target sources were observed for 1 hour in VLA configuration A, which is the one that achieves the highest C-band resolution, of $\sim$0.33 arcsec. All targets were phase-referenced with a close, bright and compact source, following a phase-target calibrator cycle of 9 minutes, 7 minutes on the target source and 2 minutes on the phase calibrator (see Table \ref{tab:host} for more details about the target sources). The flux calibrator 3C286 was observed for 6.5 minutes before the observation of each target.
    The correlator configuration was set to 32 spectral windows, each divided into 128 channels.
    
    \begin{table*}
    	\centering
    	\caption{\label{tab:host} VLA target sample of MaNGA AGN dwarf galaxy candidates.}
    	\begin{tabular}{cccccc} 
    		\hline
    		MaNGA plateifu & RA(J2000) & Dec. (J2000) & \begin{tabular}[c]{@{}c@{}}D$_\mathrm{L}$ \\ (Mpc)\end{tabular}& \begin{tabular}[c]{@{}c@{}}MaNGA \\ BPT\end{tabular} & \begin{tabular}[c]{@{}c@{}}SDSS \\ BPT\end{tabular}\\
    		\hline
    	   8442-1901 & 13:18:42.222 & +32:55:07.39 & 151.5 & AGN & Composite\\
            8255-12704 & 11:00:28.226 & +44:14:37.07 & 105.6& AGN & Star-forming\\
    	   8465-9101 & 13:10:19.259 & +47:07:26.73 & 102.1& LINER & Composite\\
          8932-9102 & 13:06:36.374 & +27:52:23.52 & 87.5& AGN & Star-forming\\
    		\hline
    	\end{tabular}
        \tablefoot{ D$_\mathrm{L}$ corresponds to the luminosity distance.}
    \end{table*}
    
     The Common Astronomy Software Applications (\textsc{casa}) package version 6.4.1.12 was used for the data reduction\footnote{https://casa.nrao.edu/}. The VLA calibration pipeline was run first. The pipeline was then rerunned after we performed a visual inspection and further data flagging. The complete data sets were imaged with the \textsc{casa} task \textsc{tclean} adopting natural weighting, which optimises the sensitivity. In addition, we produced two different images splitting the complete data set into different frequency ranges, one from 4 GHz to 6 GHz (centred at 5 GHz) and the other one from 6 GHz to 8GHz (centred at 7 GHz). We used the same task and weighting as in the complete data set imaging.
     \\
     Images showing the radio emission from galaxies 8442-1901, 8255-12794, 8465-9101 and 8932-9101 at 6 GHz are shown in Figures \ref{fig:bpt1}, \ref{fig:bpt2},\ref{fig:bpt3} and \ref{fig:bpt4}. The RMS (Root Mean Square) noise of each image has been obtained from regions far away from the source. All target sources show extended emission formed by one or more components. The different components have been selected in each image (see Figures \ref{fig:bpt1}, \ref{fig:bpt2}, \ref{fig:bpt3} and \ref{fig:bpt4}) and a Gaussian fit has been performed on these regions to obtain the size of the emission, the integrated flux and the peak emission flux. The results obtained can be read in Table \ref{tab:gausfit}. Using the maximum peak flux and the RMS noise of each image, the signal-to-noise ratio (S/N) of each region has been computed (see Table \ref{tab:gausfit}). Regions with a S/N < 5 are not considered a clear emission as the detected signal could be due to noise. The fluxes resulting from the Gaussian fits were used to compute the radio luminosity $L_\mathrm{6\,GHz}$ of the different components (see Table \ref{tab:L}). 
     \\
     Selecting the same regions as in the 6 GHz images, we applied the same procedure for the 5 GHz and the 7 GHz images to be able to compute the spectral index.

\subsection{Detecting outflows signatures}     
\label{outflow detection}
\cite{rodriguez2025manga} studied the presence of AGN outflows from the MaNGA sample of 2292 dwarf galaxies with strong AGN ionisation derived by \cite{mezcua2024manga} and which includes the four target sources here analysed. Their analysis is based on the detection of a broad component in the [OIII]$\lambda5007$ emission line in the stacked spectrum of the AGN and composite spaxels, followed by an individual analysis of the [OIII] broad component in these spaxels. They considered as dwarf galaxies with AGN outflow candidates those with at least ten consecutive spaxels with a broad component in the [OIII] line, and got a final sample of 13 dwarf galaxies with AGN outflows candidates. One of the galaxies studied in this paper, 8442-1901, presents an AGN outflow based on the \cite{rodriguez2025manga} analysis.

\section{Results and discussion}

    \begin{table*}[]
    \footnotesize
    \centering
    \caption{Results of the Gaussian fitting.}
    \begin{tabular}{cccccccc}
    \hline
    MaNGA plateifu & \begin{tabular}[c]{@{}c@{}}Clean beam size \\ (arcsec $\times$ arcsec)\end{tabular} & \begin{tabular}[c]{@{}c@{}}RMS noise \\ ($\mu$Jy beam$^{-1}$)\end{tabular} & Region & S/N          & \begin{tabular}[c]{@{}c@{}}Peak flux\\ ($\mu$Jy beam$^{-1}$)\end{tabular} & \begin{tabular}[c]{@{}c@{}}Integrated flux\\ ($\mu$Jy)\end{tabular} & \begin{tabular}[c]{@{}c@{}}Size\\ (arcsec $\times$ arcsec)\end{tabular} \\ \hline
    \\
    8442-1901                  & 0.42 $\times$ 0.37                                                        & 3.6                                                           & A      & 81.0 $\pm$ 8.0 & 291.0 $\pm$ 12.0                                                     & 500 $\pm$ 29                                                          & 0.36 $\times$ 0.31                   \\ \\
    
    8255-12704                 & 0.65 $\times$ 0.35                                                        & 4.0                                                           & A      & 13.1 $\pm$ 0.5 & 52.7 $\pm$ 0.2                                                       & 161.7 $\pm$ 0.9                                                       & 0.86 $\times$ 0.52                   \\ \\
    
    \multirow{4}{*}{8465-9101} & \multirow{4}{*}{0.53 $\times$ 0.35}                                       & \multirow{4}{*}{4.1}                                          & A      & 7.5 $\pm$ 1.0  & 30.8 $\pm$ 4.0                                                       & 51.3 $\pm$ 9.9                                                        & 0.61 $\times$ 0.51                \\
                               &                                                                         &                                                               & B      & 5.8 $\pm$ 0.5  & 23.9 $\pm$ 2.0                                                       & 58.2 $\pm$ 6.6                                                        & 1.0 $\times$ 0.46            \\
                               &                                                                         &                                                               & C      & 6.7 $\pm$ 0.4  & 27.8 $\pm$ 1.7                                                       & 81.7 $\pm$ 6.7                                                        & 1.2 $\times$ 0.47             \\
                               &                                                                         &                                                               & D      & 8.4 $\pm$ 0.6  & 34.7 $\pm$ 2.4                                                       & 90.1 $\pm$ 8.3                                                        & 0.94 $\times$ 0.51              \\ \\
                               
    \multirow{3}{*}{8932-9102} & \multirow{3}{*}{0.52 $\times$ 0.36}                                       & \multirow{3}{*}{4.0}                                          & A      & 12.2 $\pm$ 1.1 & 48.6 $\pm$ 4.1                                                       & 168.0 $\pm$ 18.0                                                      & 0.95 $\times$ 0.37       \\
                               &                                                                         &                                                               & B      & 5.3 $\pm$ 0.4  & 21.0 $\pm$ 1.6                                                       & 20.1 $\pm$ 2.7                                                       & 0.49 $\times$ 0.36                                     \\
                               &                                                                         &                                                               & C      & 5.3 $\pm$ 0.7  & 53.0 $\pm$ 10.0                                                        & 53.0 $\pm$ 10.0                                                         & 0.90 $\times$ 0.17              \\ \\\hline
    \end{tabular}
    \label{tab:gausfit}
    \tablefoot{Size corresponds to that deconvolved from the beam. Only emission regions with an S/N > 5 were considered. Each region has been marked with a letter and can be identified in Figures \ref{fig:bpt1}, \ref{fig:bpt2}, \ref{fig:bpt3} and \ref{fig:bpt4}.}
    \end{table*}
    \normalsize
 
    \begin{table*}[]
    \centering
    \caption{Radio luminosities and spectral index of the VLA sample of AGN dwarf galaxy candidates.}
    \begin{tabular}{ccccccc}
    \hline
    \begin{tabular}[c]{@{}c@{}}MaNGA\\ plateifu\end{tabular}& \begin{tabular}[c]{@{}c@{}}L$_{\nu, \mathrm{thermal}}$\\ ($10^{26}$ erg s${-1}$ Hz${-1}$)\end{tabular} & \begin{tabular}[c]{@{}c@{}}L$_{\mathrm{SNR/SN_e}}$\\ ($10^{26}$ erg s${-1}$ Hz${-1}$)\end{tabular} & \begin{tabular}[c]{@{}c@{}}L$_\mathrm{pop}$\\ ($10^{26}$ erg s${-1}$ Hz${-1}$)\end{tabular} & Regions & \begin{tabular}[c]{@{}c@{}}L$_\mathrm{6\,GHz}$\\ ($10^{26}$ erg s${-1}$ Hz${-1}$)\end{tabular} & $\alpha$ \\ \hline \\
    8441-1901                  & 23.6                                                                                       & 1.1                                                                                   & 9.9                                                                             & A       & 80.0 $\pm$ 3.2    & -0.35 $\pm$ 0.03                                                                \\ \\
    8255-12701                 & 11.4                                                                                       & 0.4                                                                                   & 3.8                                                                             & A       & 7.0 $\pm$ 0.03 & -0.5 $\pm$ 0.1                                                                \\ \\
    \multirow{4}{*}{8465-9101} & \multirow{4}{*}{2.9}                                                                      & \multirow{4}{*}{0.1}                                                                  & \multirow{4}{*}{1.0}                                                            & A       & 3.8 $\pm$ 0.5 & -0.5 $\pm$ 0.1                                                                 \\
                               &                                                                                            &                                                                                       &                                                                                 & B       & 2.9 $\pm$ 0.2 & -0.6 $\pm$ 0.1                                                     \\
                               &                                                                                            &                                                                                       &                                                                                 & C       & 3.4 $\pm$ 0.2  & -0.5 $\pm$ 0.1                                                               \\
                               &                                                                                            &                                                                                       &                                                                                 & D       & 4.3 $\pm$ 0.3  & -0.5 $\pm$ 0.1                                                               \\ \\
    \multirow{3}{*}{8932-9102} & \multirow{3}{*}{23.4}                                                                      & \multirow{3}{*}{0.9}                                                                  & \multirow{3}{*}{8.2}                                                            & A       & 4.5 $\pm$ 0.4 & -0.4 $\pm$ 0.1                                                                \\
                               &                                                                                            &                                                                                       &                                                                                 & B       & 1.9 $\pm$ 0.1  & -0.4 $\pm$ 0.1                                                             \\
                               &                                                                                            &                                                                                       &                                                                                 & C       & 1.9 $\pm$ 0.3  & -0.6 $\pm$ 0.1                                                              \\ \\ \hline
    \end{tabular}
    \label{tab:L}
    \tablefoot{L$_{\nu, \mathrm{thermal}}$, L$_{\mathrm{SNR/SN_e}}$ and L$_\mathrm{pop}$\ are the expected luminosity radio of star forming regions, individual SNR/SN$_{\mathrm{e}}$ and SNRs/SNe$_{\mathrm{e}}$ populations respectively. L$_\mathrm{6\,GHz}$ is the radio luminosity density for each region selected.  $\alpha$ is the computed spectral index for each source.}
    \end{table*}
    
    \subsection{Origin of the radio emission}

    \begin{figure*}
    \centering
    \begin{minipage}{0.48\textwidth}
        \centering
        \includegraphics[height=7.5cm]{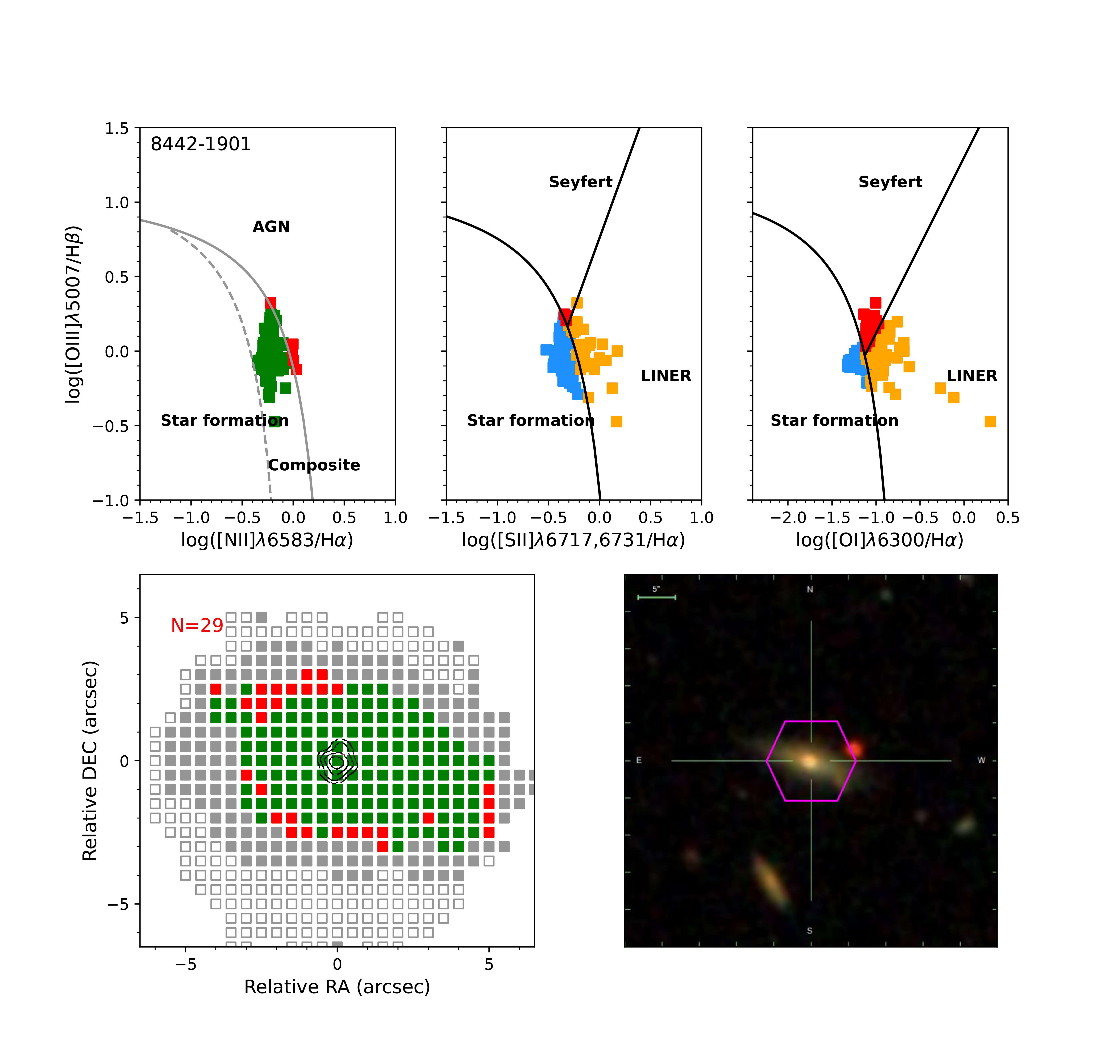}
    \end{minipage}
	\hspace{0mm}
    \begin{minipage}{0.48\textwidth}
        \centering
        \includegraphics[height=7.5cm]{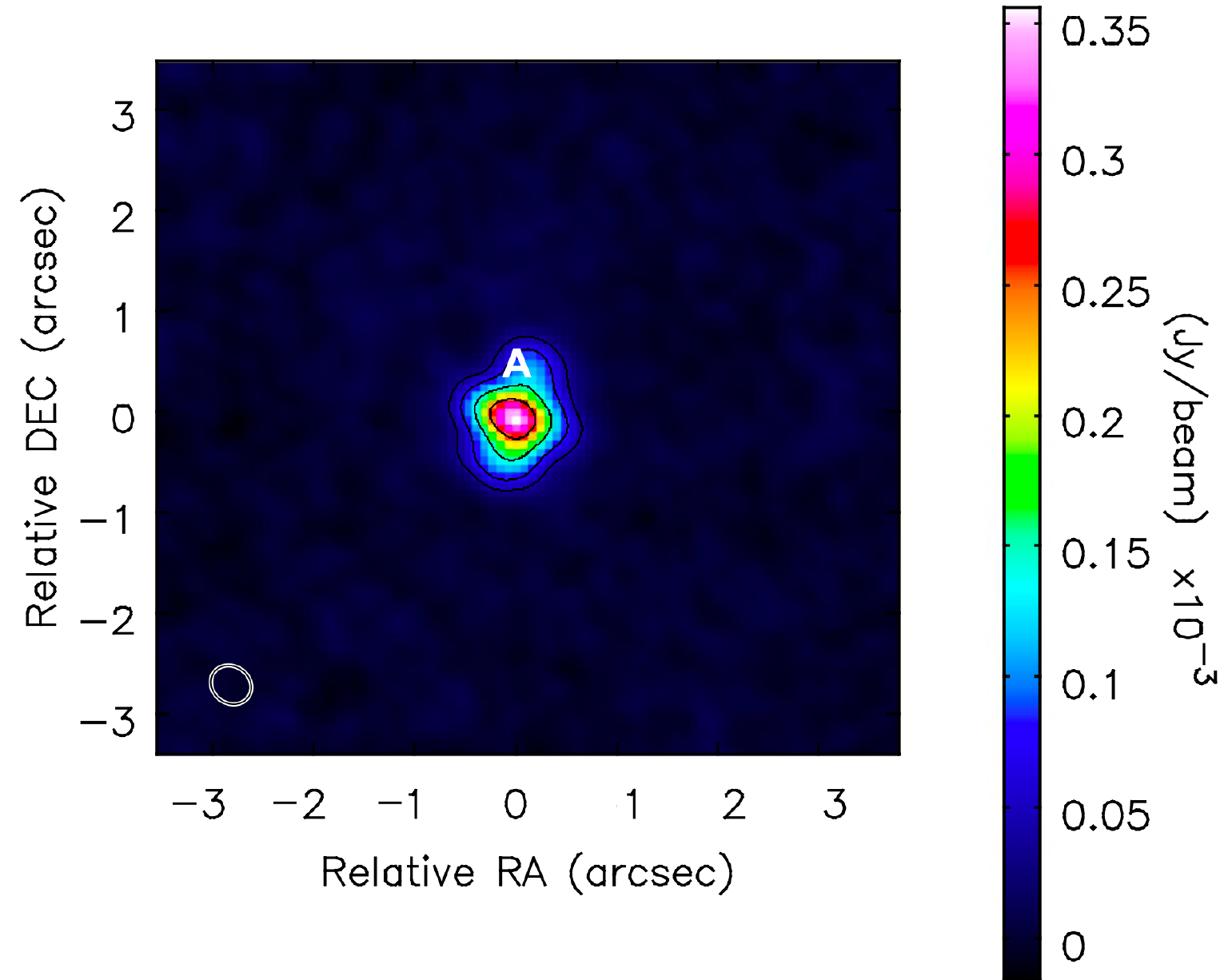}
    \end{minipage}
    \caption{Left panel: BPT diagram used to distinguish between ionisation by AGN (red spaxels), star formation (blue spaxels), composite (green spaxels) and LINER (yellow spaxels). Figure adapted from from \citet{Mezcua2020}. Middle panel: The left figure shows the spatial distribution of the BPT-classified spaxels (colour-coded as in the top left panel). Empty squares mark the IFU coverage, grey squares those spaxels with S/N > 1. The radio contours at (10, 20, 40, 70) times the off-source RMS noise have been added on the image. The right figure is the SDSS composite image. The pink hexagon shows the IFU coverage. Figure taken from \citet{Mezcua2020}. Right panel : VLA radio image at 6 GHz. The colour bar indicates the flux of each pixel in Jy beam$^{-1}$. Regions selected for Gaussian fitting are indicated by a letter. In the lower right corner is the clean beam (solid white or solid black line), which indicates the minimum beam size resolved in the observation and whose size is shown in Table \ref{tab:gausfit}. The radio contours have been added on the image.}
    \label{fig:bpt1}
    \end{figure*}

    \begin{figure*}
    \centering
    \includegraphics[width=\textwidth]{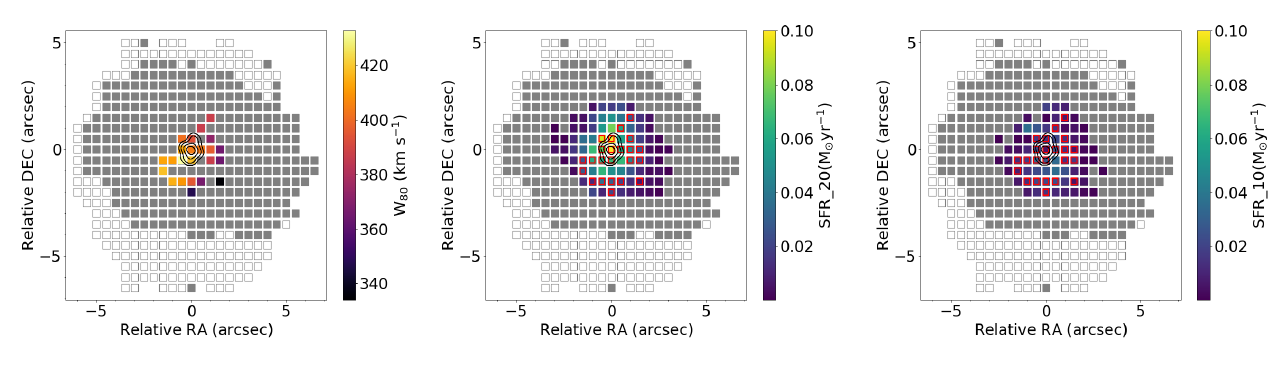} 
    \caption{Left: Spatial W$_{80}$ outflow velocity distribution of the galaxy 8442-1901. Middle: Star formation rate over the las 20 Myr. Right: SFR over the last 10 Myr. The spaxels with outflow signatures have red edges in the two SFR maps. The VLA radio contours were added in the three maps. }
    \label{figure:vel_sfr}
    \end{figure*}
    
    \begin{figure} [h!]
        \centering
        \resizebox{\hsize}{!}{\includegraphics[width = \linewidth]{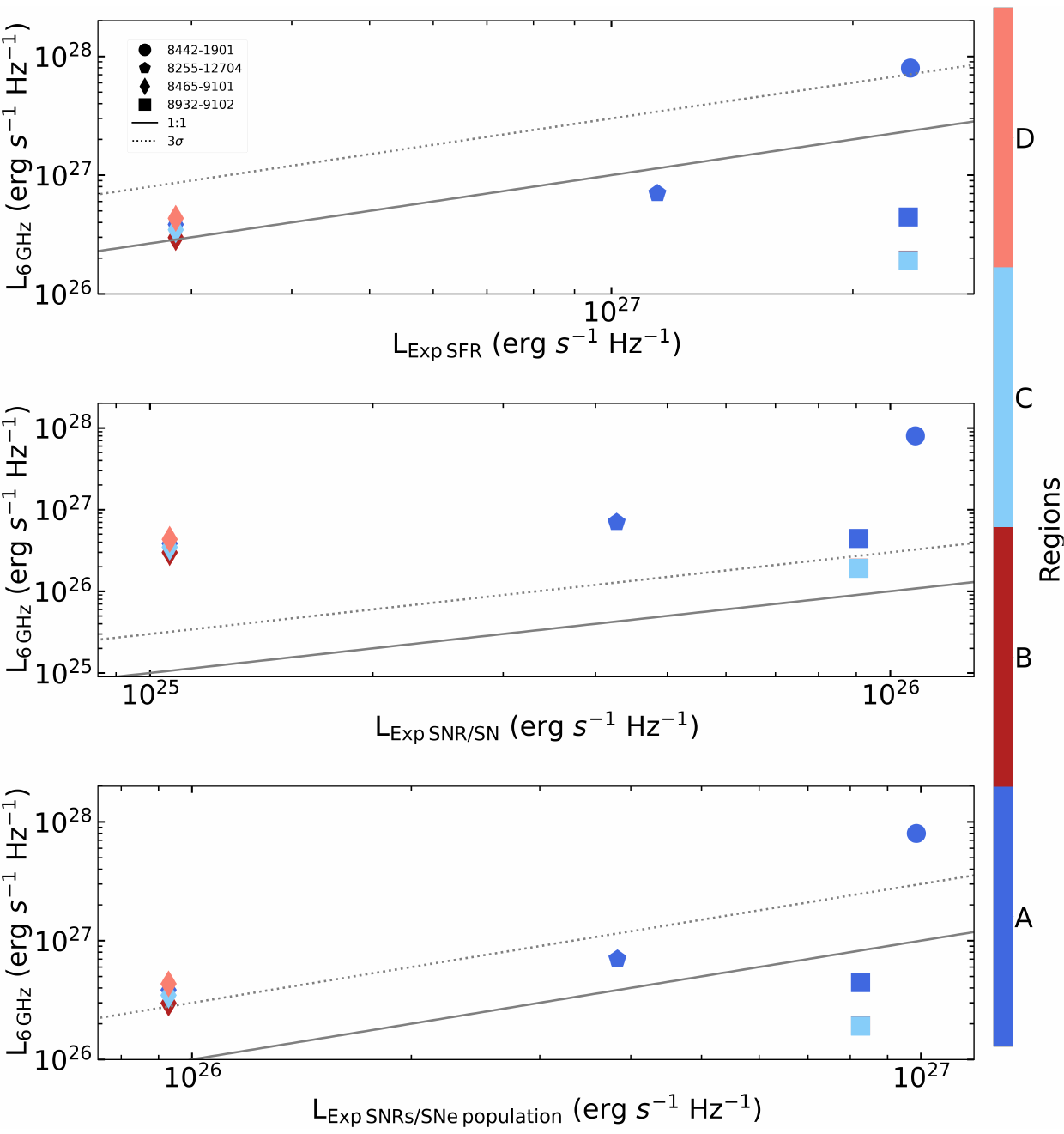}}
        \caption{Possible origin of the radio emission detected in our sample of MaNGA AGN candidates. The grey solid line represents the one-to-one relation of the expected luminosity of each possible origin. The grey dotted line represents the luminosity 3$\sigma$ above the expected luminosity for each possible origin. Unseen error bars are smaller than the size of the plotted symbols. Top: Under the assumption that the radio emission is thermal bremsstrahlung (i.e., the origin of the emission is star formation), the 6 GHz luminosity density of our sources versus the expected luminosity of the star formation has been plotted. Middle: 6 GHz luminosity densities of our sources versus the expected luminosity of individual SNR/SN. Bottom: 6 GHz luminosity densities of our sources versus the expected cumulative luminosity of the SNRs/SNe population. Sources above the 3$\sigma$ expected luminosity cannot be explained by the specific mechanism since individual regions cannot have luminosities that exceed what is expected for each mechanism in the entire galaxy. Sources below this region can have their origin in each specific mechanism.}
        \label{fig:Lvs}
    \end{figure}

    One of the aims of this project is to distinguish whether the detected radio emission comes from an AGN or has another origin. In this wavelength range we have to consider possible contamination from supernova remnants (SNRs), young supernovae (SNe) and ionised gas from HII regions. The radio emission of X-ray binaries is typically fainter ($\sim10^{17}$ W Hz${-1}$; \citeauthor{Gallo2018} \citeyear{Gallo2018}) than the detected radio emission, which is of the order of $10^{19}$ W Hz${-1}$. Therefore, they do not contaminate our radio emission.
    \\
    The first possibility is that the radio emission is produced by thermal bremsstrahlung of ionised hydrogen in star-forming regions. Under the assumption that the detected emission comes from HII regions, the $Q_{\mathrm{Lyc}}$ production ratio of the continuum Lyman photons will be estimated to calculate the expected luminosity density of these regions. The ratio $Q_{\mathrm{Lyc}}$ can be computed by the following equation \citep{Kennicutt1998}:
    \begin{equation}
        \mathrm{SFR}(M_{\odot}\mathrm{yr}^{-1}) = 1.08\times 10^{-53}Q_{\mathrm{Lyc}}(\mathrm{s^{-1}})
        \label{eq:Q}
    \end{equation}
    where SFR is the star formation rate of the host galaxy and is taken from \citet{Mezcua2024}. To calculate the expected luminosity density $L_{\nu, \mathrm{thermal}}$ of the HII regions, the following equation has been used \citep{Condon1992}:
    \begin{equation}
        \left(\frac{Q_{\mathrm{Lyc}}}{s^{-1}}\right) \gtrsim 6.3\times 10^{52}\left(\frac{T_e}{10^4\hspace{1mm}\mathrm{K}}\right)^{-0.45}\left(\frac{\nu}{\mathrm{GHz}}\right)^{0.1}\times\left(\frac{L_{\nu, \mathrm{thermal}}}{10^{20}\hspace{1mm}\mathrm{WHz}^{-1}}\right)
        \label{eq:L_sfr}
    \end{equation}
    where $\nu = $ 6 GHz and the electron temperature is  $T_e = 10^4$ K. The values obtained can be seen in Table \ref{tab:L}.
    \\
    Individual SNR or SN can also be the origin of the detected radio emission. The expected SNR/SN luminosity density can be computed as \citep{Chomiuk2009}:
    \begin{equation}
        L_{\mathrm{SNR/SN_e}} = (95^{+31}_{-23})\mathrm{SFR}^{0.98\pm 0.12}
    \end{equation}
    where $L_{\mathrm{SNR/SN_e}}$ is the luminosity density in units of $10^{24}$ erg s$^{-1}$ Hz$^{-1}$ for observations centred at 1.4 GHz and the SFR is in units of M$_{\odot}$ yr$^{-1}$.
    \\
    Finally, the radio emission may be also due to multiple SNRs/SNe. Their expected luminosity is calculated using the luminosity function:
    \begin{equation}
        n(L) = \frac{dN}{dL} = 92 \times \mathrm{SFR} \times L^{-2.07}
    \end{equation}
    where $n(L)$ is the number of SNRs/SNe with luminosity density $L$ at 1.4 GHz. The total expected luminosity has been calculated by performing the following integral:
    \begin{equation}
        L_{\mathrm{pop}} = \int n(L) L dL
    \end{equation}
    where the integral limits considered are 0.1 and $10^4$ to cover the full range of SNRs/SNe presented in \citet{Chomiuk2009}.
    \\
    The spectral index ($\alpha$) can help us to consolidate the origin of the radio emission. It is defined as the power-law spectrum:
    \begin{equation}
        S_{\nu} \propto \nu^{\alpha}
    \end{equation}
    where $S$ is the radio flux at a frequency $\nu$.
    The established classification differentiates between a steep spectrum ($\alpha \leq -0.5$), a flat spectrum ($-0.5 < \alpha \leq 0$) and an inverted spectrum ($0 < \alpha$; e.g., \citealt{Panessa2013}).
    The computed spectral index can be found in Table \ref{tab:L}.
    \\
    To be able to compare the detected $L_{\mathrm{6\,GHz}}$ with the computed expected luminosities, these last ones were scaled from 1.4 GHz to 6 GHz using the mean of the computed spectral index for each region. The results for the scaled expected luminosities can be found in Table \ref{tab:L}.

    \subsection{Jet power of an AGN}
    The radio luminosity of the AGN emission can be used to compute the total AGN's jet power ($Q_{\mathrm{jet}}$). The correlation between 1.4 GHz radio luminosity and the jet kinetic power from \citet{Cavagnolo2010} has been used to estimate the $Q_{\mathrm{jet}}$ of the AGN in our sample:
    \begin{equation}
        \mathrm{log}\hspace{1mm}Q_{\mathrm{jet}} = 0.75\hspace{1mm}(\pm 0.14)\hspace{1mm} \mathrm{log}\hspace{1mm}P_{1.4} + 1.91\hspace{1mm}(\pm 0.18)
        \label{jp}
    \end{equation}
    where $Q_{\mathrm{jet}}$ is the jet power in units of $10^{42}$ erg s$^{-1}$ and $P_{1.4}$ is the 1.4 GHz radio luminosity in units of $10^{40}$ erg s$^{-1}$. We note that this correlation has been derived for giant elliptical galaxies. Jet launching is thought to be a universal mechanism that scales with BH mass and that also holds for low-mass BHs \citep{Gultekin2014} as those expected in dwarf galaxies. We therefore assume that the \citet{Cavagnolo2010} correlation scales down to the dwarf galaxy regime, without necessarily introducing an under or overestimation in the jet power. However, given that direct measurements of the correlation in the low-mass or low-power regime are not available, we raise the flag that the values should be taken as rough estimates.
        
    \subsection{Individual targets}
    
    The highest expected radio luminosity for all galaxies in our sample comes from thermal emission (see Fig. \ref{fig:Lvs}). This means that if a source has a $L_{\mathrm{6\,GHz}}$ greater than 3$\sigma$ of the expected thermal luminosity, it could indicate the presence of an AGN.

    \subsubsection{Target source 8442-1901}
    \label{sect:8442}

    As shown in Figure \ref{fig:bpt1}, the radio emission of 8442-1901 consists of a single region (marked with an 'A') that is resolved. The radio emission $L_{\mathrm{6\,GHz}}$ is more than 3$\sigma$ higher than that of stellar processes (see Fig. \ref{fig:Lvs}), indicating an AGN origin.
    The classifications based on radio emission and the BPT diagram derived from MaNGA IFU data are in agreement: Region A is classified as composite in the BPT diagram, which indicates that the observed emission results from a combination of star formation and AGN activity. Additionally, the radio emission is concentrated in the galaxy's centre, giving support to the presence of a nuclear AGN. The computed spectral index for this radio source is flat, which is commonly associated with AGN activity. 
    \\
    The size of region A is of $\sim$265 pc and we find a kinetic jet power of $Q_{\mathrm{jet}} = 1.2 \times 10^{42}$ erg s$^{-1}$. This kinetic jet power is comparable with that of jets in massive galaxies ($Q_{\mathrm{jet}} \geq 10^{42}$ erg s$^{-1}$; e.g. \citeauthor{Merloni2007} \citeyear{Merloni2007}; \citeauthor{Mezcua2014} \citeyear{Mezcua2014}), whose jet mechanical feedback can strongly affect the star formation of the host galaxy. 
   \\
    Previous studies of jets in dwarf galaxies have shown that their powers can be comparable to those observed in massive galaxies. For example, the jet in NGC 5252 (observed with Very Long Baseline Array, VLBA; \citealt{Mezcua2018b}) is as luminous as that in 8442-1901 and the jet power of SDSS J090613.77+561015.2 (Observed with EVN; \citealt{Yang2020}) remains within the same order of magnitude as in 8442-1901. The jet power of the sample of dwarf galaxies hosting radio AGN in \citet{Davis2022} is also of the same order of magnitude as in 8442-1901.
    \\
    As described in Sect \ref{outflow detection}, 8442-1901 shows an AGN outflow (\citealt{rodriguez2025manga}). The velocity of the outflow was derived using the W$_{80}$ estimator, which measures the velocity width that contains 80$\%$ of the [OIII] flux. The AGN outflow velocity is W$_{80}=396$ km s$^{-1}$ and is consistent with escaping the dark matter halo. The kinetic energy rate of the outflow is of dE/dt$=10^{38.32}$ erg s$^{-1}$. In terms of energy, the jet can be considered the main driver of the outflow, as the jet power is four orders of magnitude higher than the kinetic energy rate of the outflow.
    \\
In Fig. \ref{figure:vel_sfr} , we show the velocity distribution of the outflow, in addition to the SFR over the last 20 and 10 Myr. For the SFR maps, we use MEGACUBES from the Laborat\'{o}rio Interinstitucional de e-Astronomia MaNGA Portal\footnote{\url{https://manga.if.ufrgs.br/}} (LIneA; \citealt{riffel2023mapping}). In these maps, a decrease of the SFR is observed in the central region where the radio emission and the AGN outflow is located. This can be explained by the capability of the outflow to escape from the gravitational bound of the dark matter halo, being able to remove gas from the central region and suppressing the star formation. This suggests the presence of negative AGN feedback in this galaxy.
\\
Jet feedback can both trigger and quench star formation. The latter could provide an alternative to environment being the driver of the red dwarf galaxy population \citep{Kaviraj2025}. The blue colours of 8442-1901 however suggest that it is not quenched by jet feedback.

    \subsubsection{Target source 8255-12704}
     As shown in Figure \ref{fig:bpt2}, the radio emission of 8255-12704 is extended and the region under study (region 'A') is resolved. We find that the emission could be due to thermal radiation or the emission from a population of SNRs/SNe, as both scenarios yield a $L_{\mathrm{6\,GHz}}$ consistent with that detected.
     
     However, if the emission was due to a single SNR/SN, the $L_{\mathrm{6\,GHz}}$ of 8255-12704 would be higher than what is expected from this type of source.
    The MaNGA IFU BPT diagram classifies Region A as a star-forming region, a result that aligns with the radio emission classification. 
    The spectral index for this region is -0.5, which is consistent with emission from a SNR/SN.
    
    \subsubsection{Target source 8465-9101}
    As shown in Figure \ref{fig:bpt3}, the radio emission of 8465-9101 is extended and Regions A, B, C and D are resolved (see Table \ref{tab:gausfit}).
    \\
    In the case of Regions A, C, and D, their $L_{\mathrm{6\,GHz}}$ is consistent with the expected thermal luminosity, suggesting that their emission is likely due to star-forming regions. The $L_{\mathrm{6\,GHz}}$ of these regions is higher than what would be expected from either a single SNR/SN or a population of SNRs/SNe. Region B, on the other hand, could have emission due to thermal radiation since its $L_{\mathrm{6\,GHz}}$ matches the expected thermal luminosity. However, the emission could also be due to a SNRs/SNe population, as its $L_{\mathrm{6\,GHz}}$ falls within a range consistent with this mechanism (see Fig. \ref{fig:Lvs}).
     Based on MaNGA IFU, Regions A, B, and C are classified as composite in the BPT diagram, while Region C is classified as a star-forming region. These optical classifications are consistent with the radio emission categorisation discussed above.  The spectral indices for Regions A, C, and D are all $\alpha$ = -0.5, indicative of emission from SNR/SN. Region B has $\alpha$ = -0.6 also consistent with SNR/SN emission.
    
    \subsubsection{Target source 8932-9102}
    As shown in Figure \ref{fig:bpt4}, the radio emission of 8932-9101 is extended. Regions A, B and C are resolved (see Table \ref{tab:gausfit}).
    For Region A the emission may be due to a population of SNRs/SNe as the $L_{\mathrm{6\,GHz}}$ is consistent with the expected emission from such sources (see Fig. \ref{fig:Lvs}). The same reasoning applies to Region B and C, but in this case, the emission is more likely from a single SNR/SN.
    In the spatially resolved BPT diagram all the regions are consistent with star-formation ionisation, which aligns with the radio emission classification. 
    The spectral indices for Regions A and B are consistent with star-forming regions, while the index for Region C aligns with that of a SNR/SN.

\section{Conclusion}

    AGN in dwarf galaxies can be identified through radio observations if they are able to produce kinetic jets detectable in the radio regime \citet{Reines2020, Yang2023, Wu2024}. If these jets are sufficiently powerful, they can impact several properties of the host galaxy, for example, by either suppressing or triggering star formation.
    \\
    In this paper we used VLA radio observations to investigate the presence of radio jets in a sample of four dwarf galaxies candidate to host AGN based on optical MaNGA IFU observations and with FIRST radio counterparts (\citealt{Mezcua2020,Mezcua2024}).
    We successfully resolve the FIRST radio emission into multiple regions and derive their radio luminosity. The radio luminosity of the different regions are compared to the expected luminosities from star-forming regions, individual SNR/SN, and SNRs/SNe populations in each host galaxy. We found that target 8442-1901 hosts an AGN. The radio emission produced by the different regions of the other three targets is of stellar origin.
    \\
    The classification based on the spatially resolved BPT diagrams derived from MaNGA IFU spectroscopy is in agreement with the radio emission results for all sources when examining specific regions.
    \\
    We also calculate the jet power for target 8442-1901, obtaining $Q_{\mathrm{jet}} \sim 10^{42}$ erg s$^{-1}$. This value suggests that dwarf galaxies can produce jets as powerful as those in massive radio galaxies as is shown in several studies (e.g, \citealt{Wrobel2006}, \citealt{Mezcua2015}, \citealt{Mezcua2018b}, \citealt{Yang2020}, \citealt{Davis2022}). The MaNGA IFU reveals that 8442-1901 also has an AGN outflow that is able to escape the dark matter halo gravitational bound (\citealt{rodriguez2025manga}). A decrease of the SFR in the central region of this galaxy is observed, suggesting the presence of negative AGN feedback and that AGN in dwarf galaxies may regulate star formation similarly to the more massive galaxies.
    
\begin{acknowledgements}
       IF acknowledges support from the Spanish Ministry of Science, Innovation and Universities through the project PRE2022-104899. MM acknowledges support from the Spanish Ministry of Science and Innovation through the project PID2021- 124243NB-C22 and financial support from AGAUR under project 2021-SGR-01270. This work was partially supported by the program Unidad de Excelencia Mar\'ia de Maeztu CEX2020-001058-M. VRM acknowledges support from the Spanish Ministry of Science, Innovation and Universities through the project PRE2022-104649. 
\end{acknowledgements}

\bibliographystyle{aa}
\bibliography{example}

\begin{appendix}
\onecolumn
\section{BPT diagram, ionisation map and radio emission}
\label{app:A}

\begin{figure*}[h]
\centering
\begin{minipage}{7.5cm}
	\includegraphics[height=7.5cm]{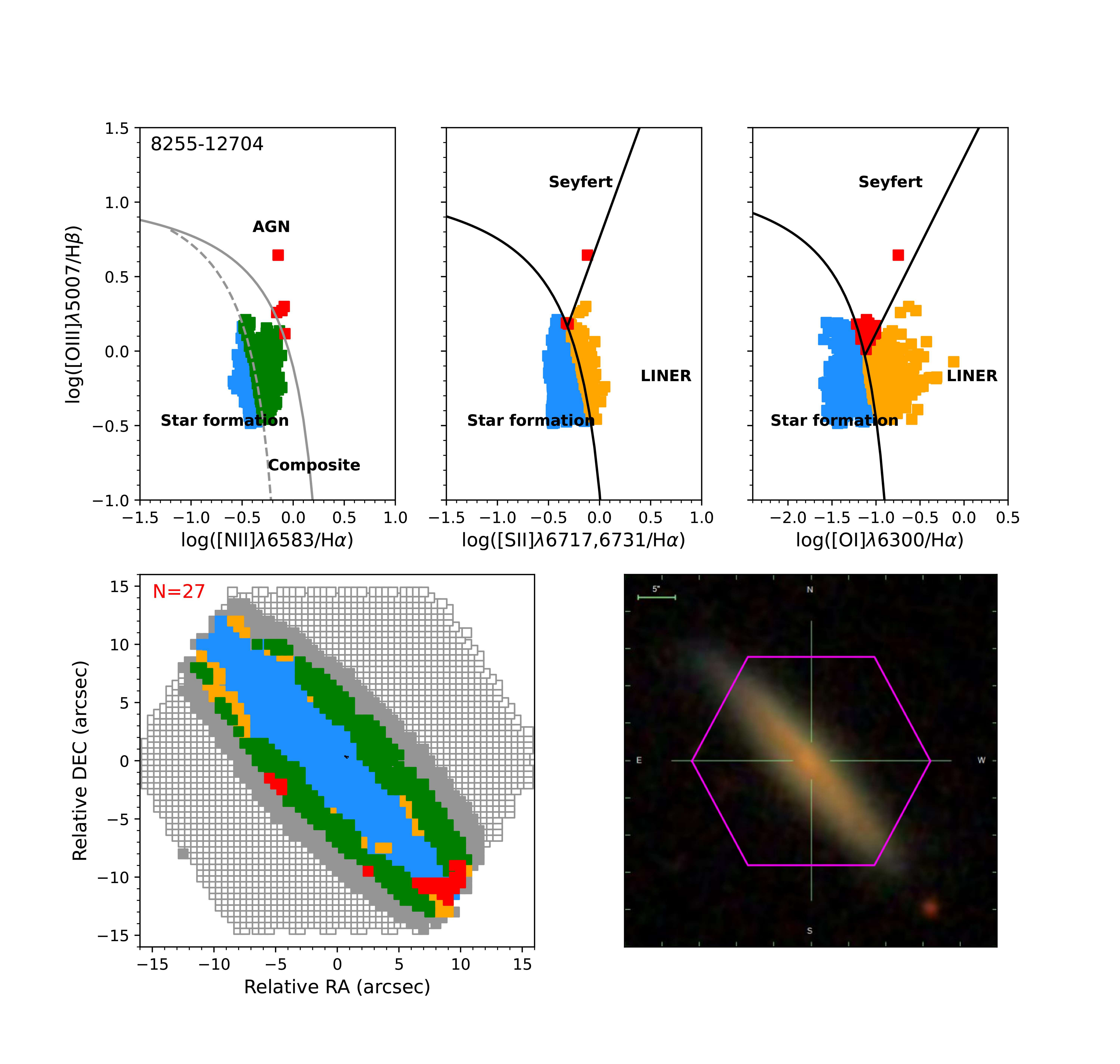}
\end{minipage}%
\begin{minipage}{7.5cm}
	\includegraphics[height=7.5cm]{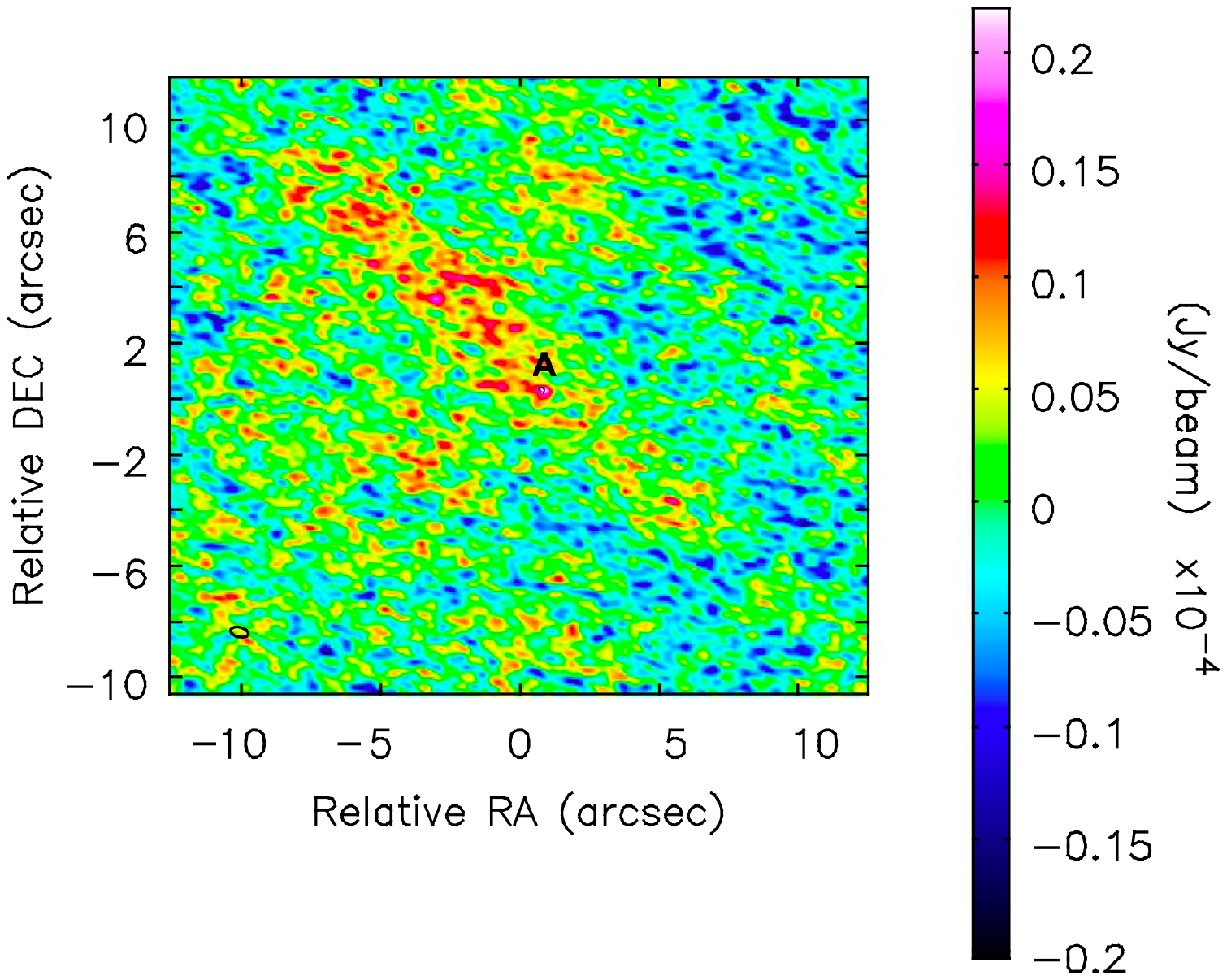}
\end{minipage}
\caption{Same caption as in figure \ref{fig:bpt1}. The VLA contours are 5 times the off-source RMS noise.}
\label{fig:bpt2}
\end{figure*}

\begin{figure*}[h]
\centering
\begin{minipage}{7.5cm}
\centering
\includegraphics[height=7.5cm]{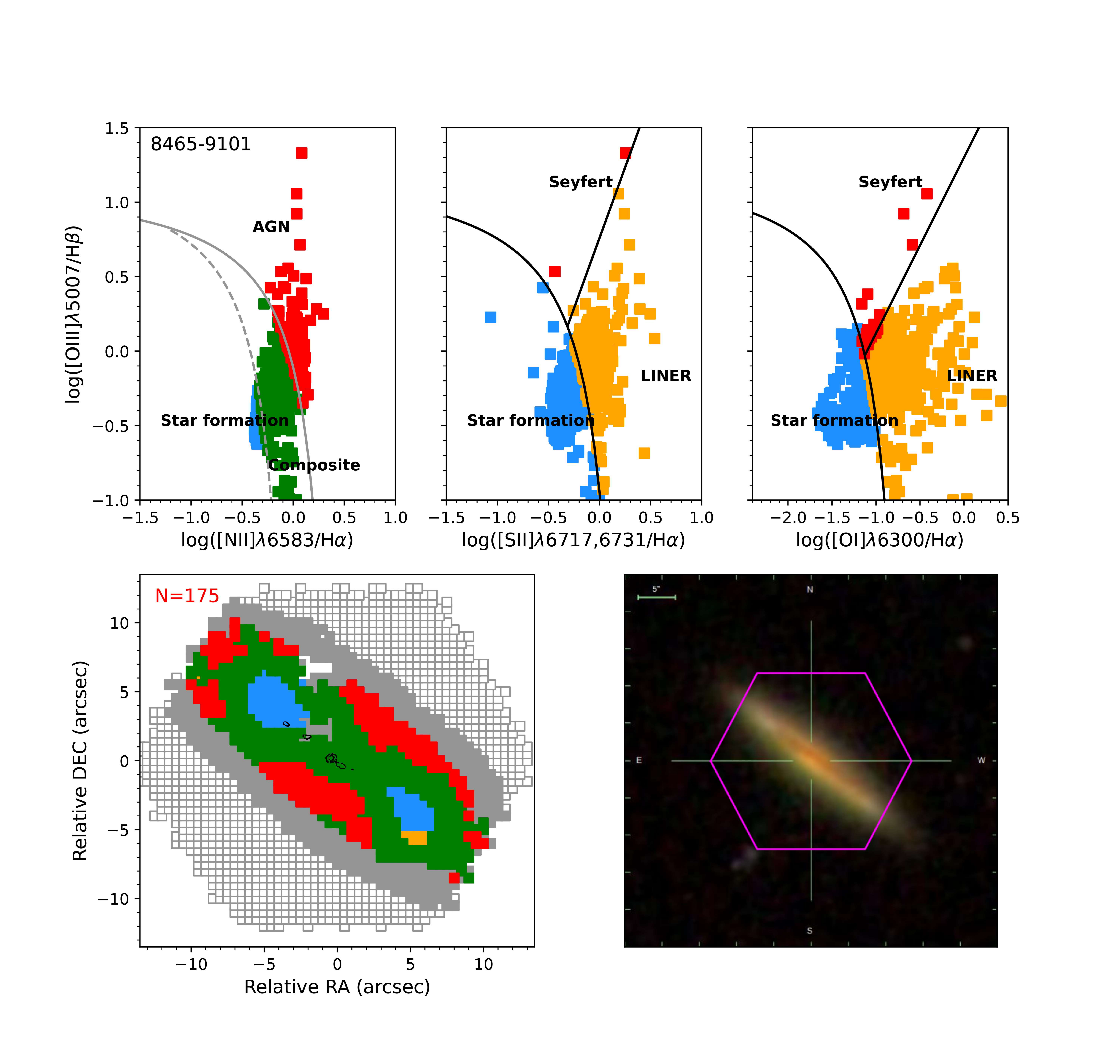}
\end{minipage}%
\begin{minipage}{7.5cm}
\centering
\includegraphics[height=7.5cm]{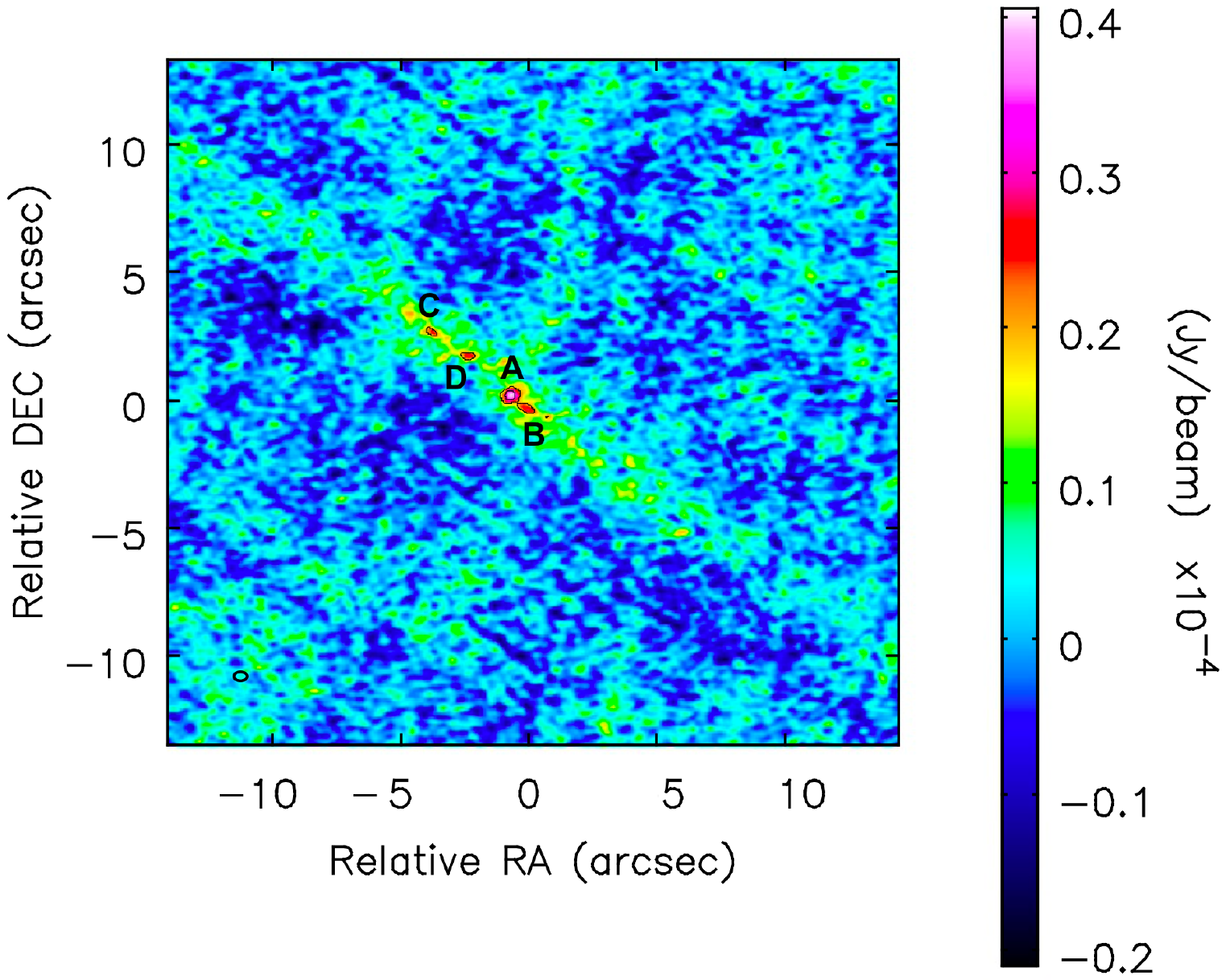}
\end{minipage}
\caption{Same caption as in figure \ref{fig:bpt1}. The VLA contours are (5, 8) times the off-source RMS noise.}
\label{fig:bpt3}
\end{figure*}

\begin{figure*}[h]
\centering
\begin{minipage}{7.5cm}
\centering
\includegraphics[height=7.5cm]{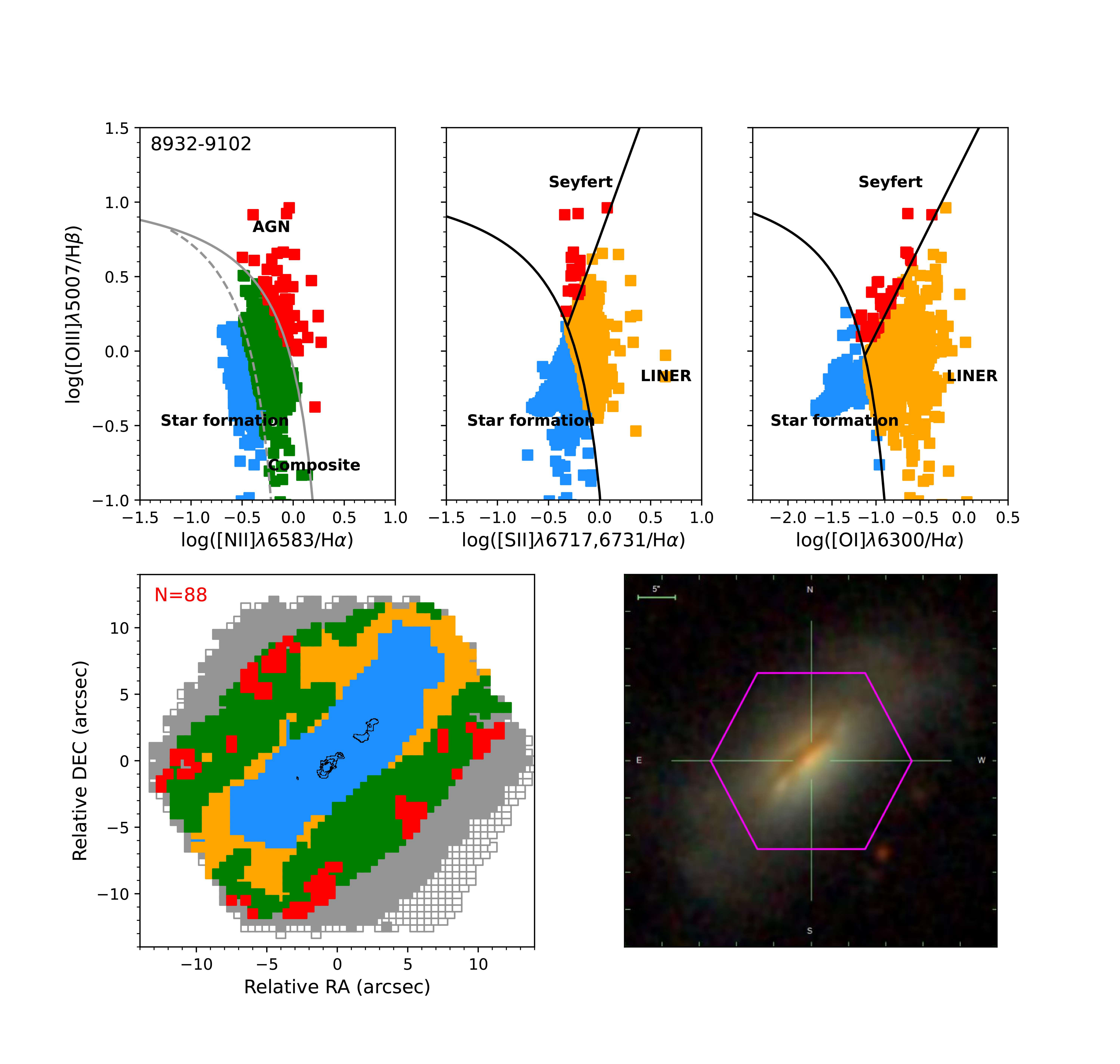}
\end{minipage}%
\begin{minipage}{7.5cm}
\centering
\includegraphics[height=7.5cm]{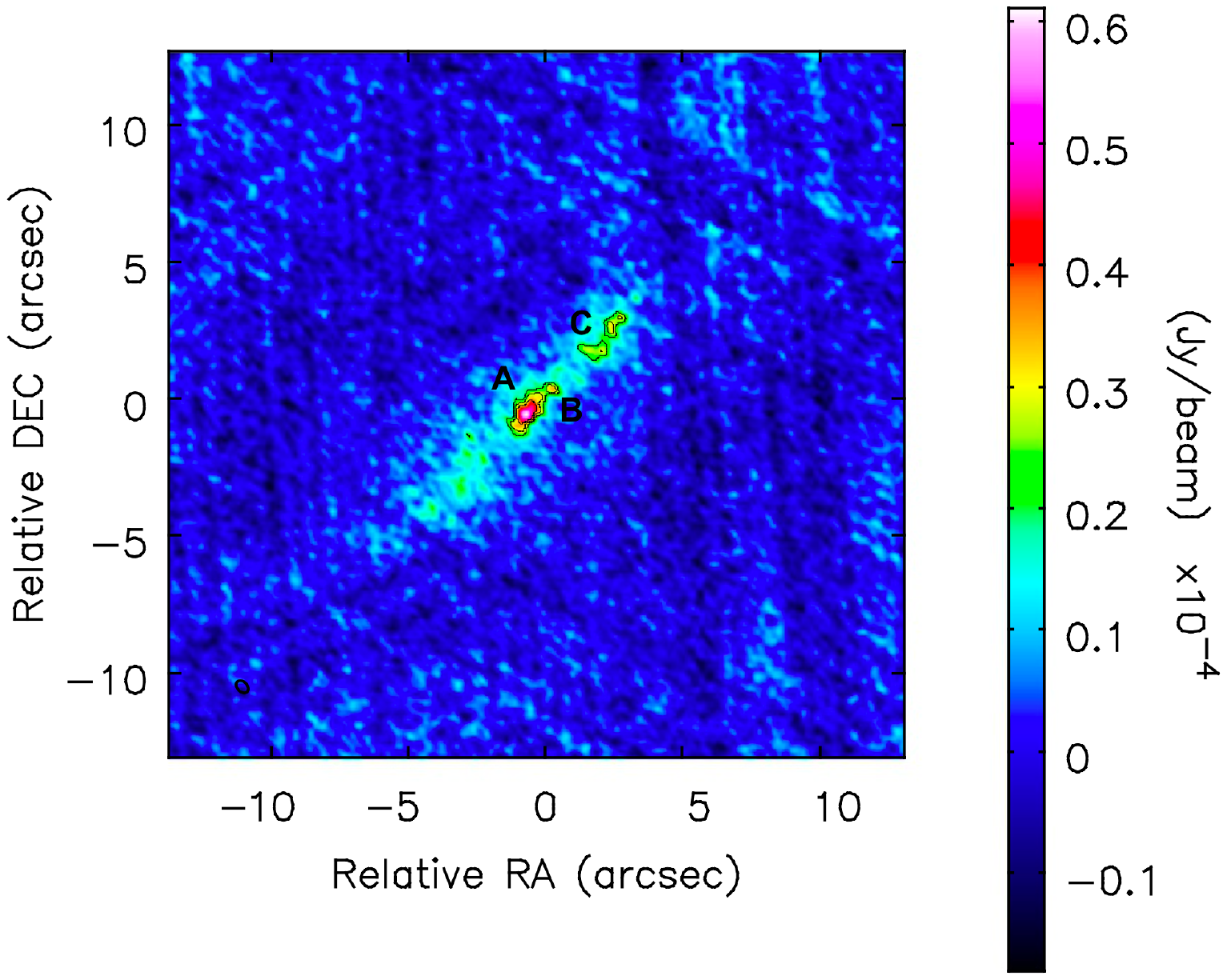}
\end{minipage}
\caption{Same caption as in figure \ref{fig:bpt1}. The VLA contours are (5, 7, 9) times the off-source RMS noise.}
\label{fig:bpt4}
\end{figure*}

Figures \ref{fig:bpt2}-\ref{fig:bpt4} show the BPT diagrams, the ionisation maps and the radio emission of the galaxies 8255-12704, 8465-9101 and 8932-1901 as in Fig. \ref{fig:bpt1}. 

\end{appendix}
\end{document}